\documentclass[%
 reprint,
superscriptaddress,
bibnotes,
 amsmath,amssymb,
 aps,
prb,
]{revtex4-2}

\usepackage{graphicx}
\usepackage{dcolumn}
\usepackage{bm}
\usepackage{caption}
\usepackage{float}
\usepackage{amsmath}
\usepackage{animate}
\usepackage{hyperref}
\usepackage{chemformula}

\pdfoutput=1
\DeclareCaptionFormat{myformat}{\hspace{1.5em}#1#2#3\par}
\begin{document}
 

\title{Ferromagnetic half levitation of LK-99-like synthetic samples}

\author{Kaizhen Guo}
\affiliation{
International Center for Quantum Materials, School of Physics, Peking University, Beijing 100871, China
}%

\author{Yuan Li}
\homepage{yuan.li@pku.edu.cn}
\affiliation{
	International Center for Quantum Materials, School of Physics, Peking University, Beijing 100871, China
}%
\affiliation{
    Collaborative Innovation Center for Quantum Matter, Beijing 100871, China
}%

\author{Shuang Jia}
\homepage{gwljiashuang@pku.edu.cn}
\affiliation{
International Center for Quantum Materials, School of Physics, Peking University, Beijing 100871, China
}%
\affiliation{
 Interdisciplinary Institute of Light-Element Quantum Materials and Research Center for Light-Element Advanced Materials, Peking University, Beijing 100871, China
}%
\affiliation{
CAS Center for Excellence in Topological Quantum Computation,
University of Chinese Academy of Sciences, Beijing 100190, China
}%

\date{\today}

\begin{abstract}

We successfully synthesized polycrystalline LK-99-like ceramic samples with a solid-state-sintering method. Powder X-ray diffraction shows that the main contents are $\mathrm{Pb_{10-x}Cu_x(PO_4)_6O}$ and $\mathrm{Cu_2S}$, consistent with recent reports [arXiv:2307.12037; arXiv:2308.01192].
In some small flaky fragments, we successfully observed ``half levitation'' atop a $\mathrm{Nd_2Fe_{14}B}$ magnet.
Using magnetization measurements on such small pieces, as well as on a large piece which does not exhibit the half levitation, we show that the samples ubiquitously contain weak yet definitive soft ferromagnetic components. We argue that, together with the pronounced shape anisotropy of the small fragments, the soft ferromagnetism is sufficient to explain the observed half levitation in strong vertical magnetic fields. Our measurements do not indicate the presence of the Meissner effect, nor zero resistance, in our samples, leading us to believe that our samples do not exhibit superconductivity. 
The precise chemical composition and the physics behind the ferromagnetic component remain outstanding questions to be addressed in future research.

\end{abstract}


\maketitle




\section{INTRODUCTION}

Since the discovery of superconductivity in mercury by Onnes in 1911, researchers have explored the remarkable phenomena for over a century. Discovery of new superconducting materials and improvement of superconducting temperature have been the top goals pursued. Recent reports of research progress include nickel-based superconductors \cite{Sun2023, zhang2023hightemperature} and nitrogen-doped lutetium hydride materials \cite{Dasenbrock-Gammon2023, Ming2023}. However, both of these systems were found to possess signatures of superconductivity only under pressure, and they required relatively complicated synthesis and preparation.
Very recently, Lee \textit{et al.} claimed \cite{Lee01,Lee02,Lee03} to have found superconductivity at room temperature and atmospheric pressure in a copper-substituted lead phosphate apatite [chemical formula: $\mathrm{Pb_{10-x}Cu_x(PO_4)_6O}$, also known as ``LK-99''].
Unlike the above two materials, the reported preparation process of LK-99 was very simple. This has led to a flurry of attempts to reproduce the experimental results \cite{india,beihang,dongnan,huake} and attain theoretical understanding of this class of materials \cite{baskaran2023broad,cabezasescares2023theoretical,griffin2023origin,kurleto2023pbapatite,lai2023firstprinciples,si2023electronic,tavakol2023minimal}.

A pivotal experimental indication reported so far has been the observation of half levitation of samples \cite{Lee01,Lee02,Lee03} in strong vertical magnetic fields generated by permanent magnets in ambient condition. Notably, such observation has been successfully reproduced by independent researchers \cite{huake}. While the observation suggests the presence of room-temperature Meissner effect in certain parts of the levitated samples, the levitation could only be considered ``half'' because parts of the samples were still in contact with the supporting surface. Indeed, two of the most important properties of a superconductor, the Meissner effect and zero resistance, have not been sufficiently demonstrated and reproduced in quantitative measurements, which adds uncertainty to the verification of LK-99 as a true room-temperature superconductor.
Although some magnetization measurements have suggested that samples may be diamagnetic in small magnetic fields (10~Oe) \cite{Lee01,Lee02,Lee03,huake}, the increase of this diamagnetic response with increasing fields (to the extent of levitation against gravity) has not been demonstrated. It should be noted that diamagnetism, which is a common presence in many insulators, is inequivalent to the Meissner effect. In some ferromagnetic systems, when the direction of the external magnetic field is opposite to the direction of the internal magnetization of the material, a seemingly diamagnetic signal may be detected under a small magnetic field.

In this work, we successfully synthesized polycrystalline LK-99-like samples and observed magnetic half-levitation phenomena in some small pieces. Using powder X-ray diffraction and energy-dispersive X-ray spectroscopy, we verified that our sample compositions are in agreement with earlier studies. Then we conducted magnetization and resistance measurements of our as-grown samples and analyzed their properties without any presumption.

Our magnetization measurements on different pieces show a ubiquitous existence of soft-ferromagnetic components, characterized by a small magnetic-hysteresis loop in field-dependent magnetization. Specifically, the curve of field-dependent magnetization measured at 100~K on a large piece, which does not show the half levitation, can be viewed as a superposition of a ferromagnetic and a linearly-diamagnetic behavior. Yet, in a small and thin sample that does show half-levitation, while we still observe small diamagnetic signals at 10~Oe, consistent with previous reports \cite{Lee01,Lee02,Lee03,huake}, the diamagnetic response is quickly overwhelmed by the ferromagnetic response at higher fields. Based on this observation and an analysis of the shape anisotropy associated with the small sample, we attribute the half levitation to the ferromagnetic response of the sample to strong external fields.

Moreover, the resistance of our samples does not show any signatures of superconductivity as reported in previous works \cite{Lee01,Lee02,Lee03,dongnan}. Instead, we observed a semiconducting behavior consistent with other reports \cite{beihang,india}. Given our observation of the ferromagnetic components and the semiconducting property, we believe that the presence of room-temperature superconductivity in LK-99 remains to be stringently verified. In particular, we caution that observation of half levitation of anisotropically-shaped samples should be interpreted with scrutiny against ferromagnetism. In addition to the pursuit of superconductivity, the nature of room-temperature ferromagnetism in such a Pb-Cu-P-O system may warrant further understanding by physicists.



\section{EXPERIMENT}

Polycrystals of $\mathrm{Pb_{10-x}Cu_x(PO_4)_6O}$~($0.9 < x < 1.1$) were grown by a solid-state-sintering method reported by Lee \textit{et al.} \cite{Lee02}. The products of the synthesis were black thick pieces with a diameter of 6~mm and a thickness of 3~mm [Fig.~\ref{figs1}(a)]. 
The surface of the sample had a layer of red transparent crystals, which did not match the appearance of the LK-99 sample reported by Lee \textit{et al.} \cite{Lee01,Lee02,Lee03}. We suspected that this may be some kind of mixed byproduct of the reaction. We then carefully removed these transparent crystals with non-magnetic tweezers before taking further measurements.
The main components were confirmed by performing powder X-ray diffraction (PXRD) measurements at room temperature using a Rigaku Mini-flux 600 instrument. The composition of our samples was verified by performing energy-dispersive X-ray spectroscopy (EDS) measurements using an FEI Helios NanoLab 600i DualBeam System.

In order to obtain samples for subsequent resistance and magnetization measurements, we remove surface impurities as much as possible from the growth products and press them into pellets again, and cut the pellets into sizes suitable for measurement.
To examine whether the samples show the magnetic-levitation characteristics as described in previous reports, we put the samples on a piece of waxed weighing paper, which was fixed stably above an empty space. Then we approach the bottom of the weighing paper with a $\mathrm{Nd_2Fe_{14}B}$ magnet with caution, avoiding direct contact between them. While large samples were all stationary, we surprisingly found that a few small particles were shaken by the movement of the approaching magnet, and that one or two flaky samples were able to spontaneously turn into a vertical position, hence showing the half levitation characteristics. Two snapshots and an illustrative drawing of such a half levitated sample are shown in the Supplementary Materials, see Fig.~\ref{figs1}(c), Fig.~\ref{figs3}, and Fig.~\ref{figs4}. We managed to isolate some of such half-levitated pieces for further investigations.

Our resistance measurement was carried out using a Quantum Design Physical Properties Measurement System (PPMS) using a standard four-probe method.

Magnetization measurements were carried out using a Quantum Design Magnetic Properties Measurement Systems (MPMS-3). Samples were loaded on a quartz holder using GE-varnish glue, which provides a negligible diamagnetic signal.

\section{RESULTS}

\begin{figure}[!htbp]
	\centering
	\includegraphics[width=\linewidth]{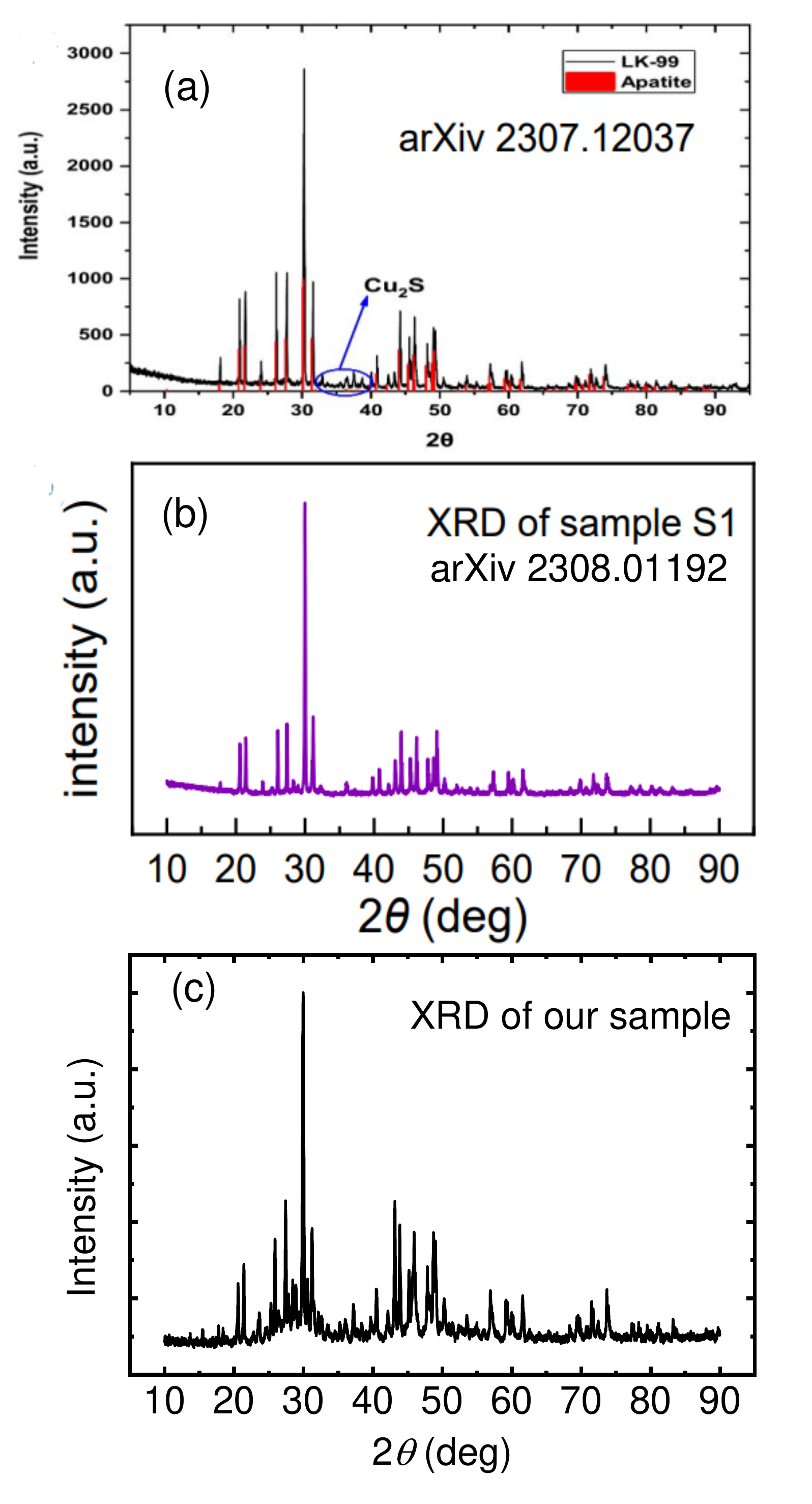}
	\caption{\label{fig1} X-ray diffraction patterns of (a) ref \cite{Lee02,Lee03} ,(b) ref \cite{dongnan} and (c) our sample.}
\end{figure}

Figure~\ref{fig1}(c) shows the PXRD pattern of our sample. Compared with those reported by references \cite{Lee02,Lee03} [Fig.\ref{fig1}(a)] and reference \cite{dongnan} [Fig.\ref{fig1}(b)], the position of the peaks is almost identical and the existence of $\mathrm{Cu_2S}$ peaks is also seen.
To further verify the elemental composition, we perform EDS measurements on our samples. The presence of lead, phosphorus, copper, oxygen, and sulfur can be detected on the surface of the sample, as seen in Fig.~\ref{figs1}(b).

\begin{figure}[!htbp]
	\centering
	\includegraphics[width=\linewidth]{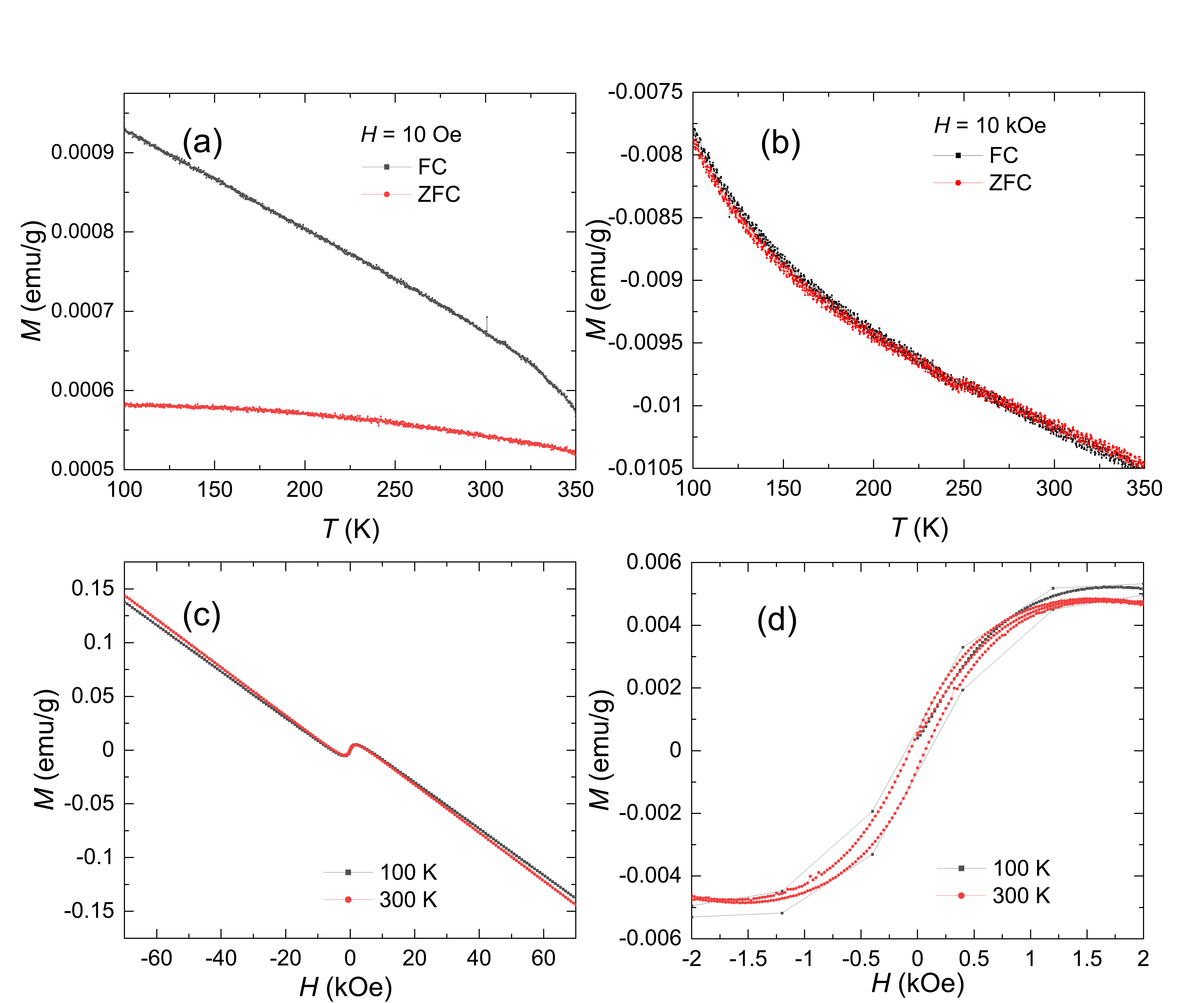}
	\caption{\label{fig2} The magnetization of sample $\mathrm{S1}$. (a) The temperature-dependent magnetization (\textit{M-T}) in a magnetic field of 10~Oe. (b) \textit{M-T} in a magnetic field of 10~kOe. (c) Field-dependent magnetization (\textit{M-H}) at 100 and 300~K. (d) Expanded view of the \textit{M-H} curves ranging from -2~kOe to 2~kOe. }
\end{figure}

First, we measured the magnetization of sample $\mathrm{S1}$ that was not half-levitated on a $\mathrm{Nd_2Fe_{14}B}$ magnet. 
Field-cooling (FC) and zero-field-cooling (ZFC) measurements were carried out successively. When the external magnetic field is 10~Oe (the same with references \cite{Lee01,Lee02,Lee03,huake}), both the FC and ZFC curves of magnetization versus temperature (\textit{M-T}) show positive magnetic moments and significant branching as illustrated in Fig.\ref{fig2}(a). 
Upon increasing the magnetic field to 10~kOe, the FC and ZFC \textit{M-T} curves remained positive and coincided, as depicted in Fig.\ref{fig2}(b).
Branching patterns in FC and ZFC curves typically manifest in ferromagnetic materials, spin glass materials, and superconductors. However, spin-glass states are more common at lower temperatures, effectively freezing the magnetic moment, and superconducting states typically yield significantly negative ZFC-magnetization values.

Consequently, this observation prompted us to acknowledge the presence of a ferromagnetic component for the first time. To further explore this, we conducted field-dependent magnetization measurements at both 100~K and 300~K, as seen in Fig.\ref{fig2}(c). The external magnetic field was increased from 0 to 70~kOe, subsequently decreased from 70 to -70~kOe, and finally increased again from -70 to 70~kOe. Comparable behaviors were observed at both temperatures.
When the magnetic field increases from 0 to 1500~Oe, the magnetization increases with the magnetic field, and then the magnetization decreases almost linearly with the magnetic field increases and even becomes negative. 
This phenomenon indicates a substantial presence of the insulating component within our sample $\mathrm{S1}$. Upon closer examination of the low-field data, a distinct magnetic-hysteresis loop emerged [Fig.\ref{fig2}(d)], further confirming the existence of a ferromagnetic phase.

\begin{figure}[!htbp]
	\centering
	\includegraphics[width=\linewidth]{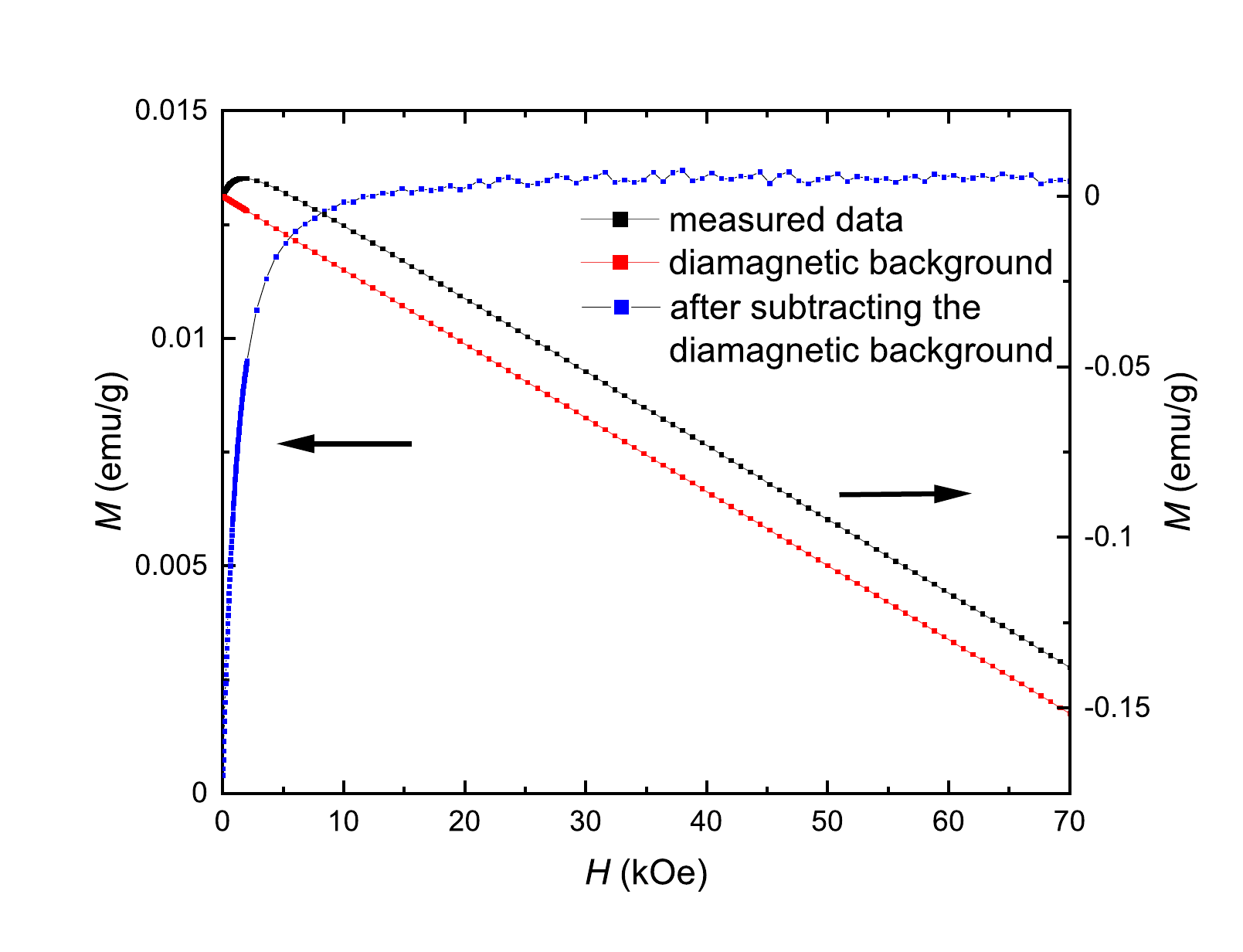}
	\caption{\label{fig3} The field-dependent magnetization of sample $\mathrm{S1}$ at 100~K (black curve). After subtracting a linear-diamagnetic part (red curve), the remainder exhibits a typical saturation above 20~kOe (blue curve). }
\end{figure}

We try to simply subtract a linear-diamagnetic part of \textit{M-H} from the measured data, taking 100~K as an example [Fig.\ref{fig3}]. After subtracting the diamagnetic background, the remainder exhibits a typical saturation above 20~kOe. Let's compare some diamagnetic materials with our sample $\mathrm{S1}$. The subtracted diamagnetic susceptibility ($\mathrm{\sim-2\times10^{-6}~emu/g}$) is larger than that of Bismuth ($\mathrm{\sim-1.6\times10^{-6}~emu/g}$) and water ($\mathrm{\sim-10^{-7}~emu/g}$) but smaller than that of Pyrolytic carbon ($\mathrm{\sim-4\times10^{-6}~emu/g}$), indicating that this part is not from the superconductivity.

\begin{figure}[!htbp]
	\centering
	\includegraphics[width=\linewidth]{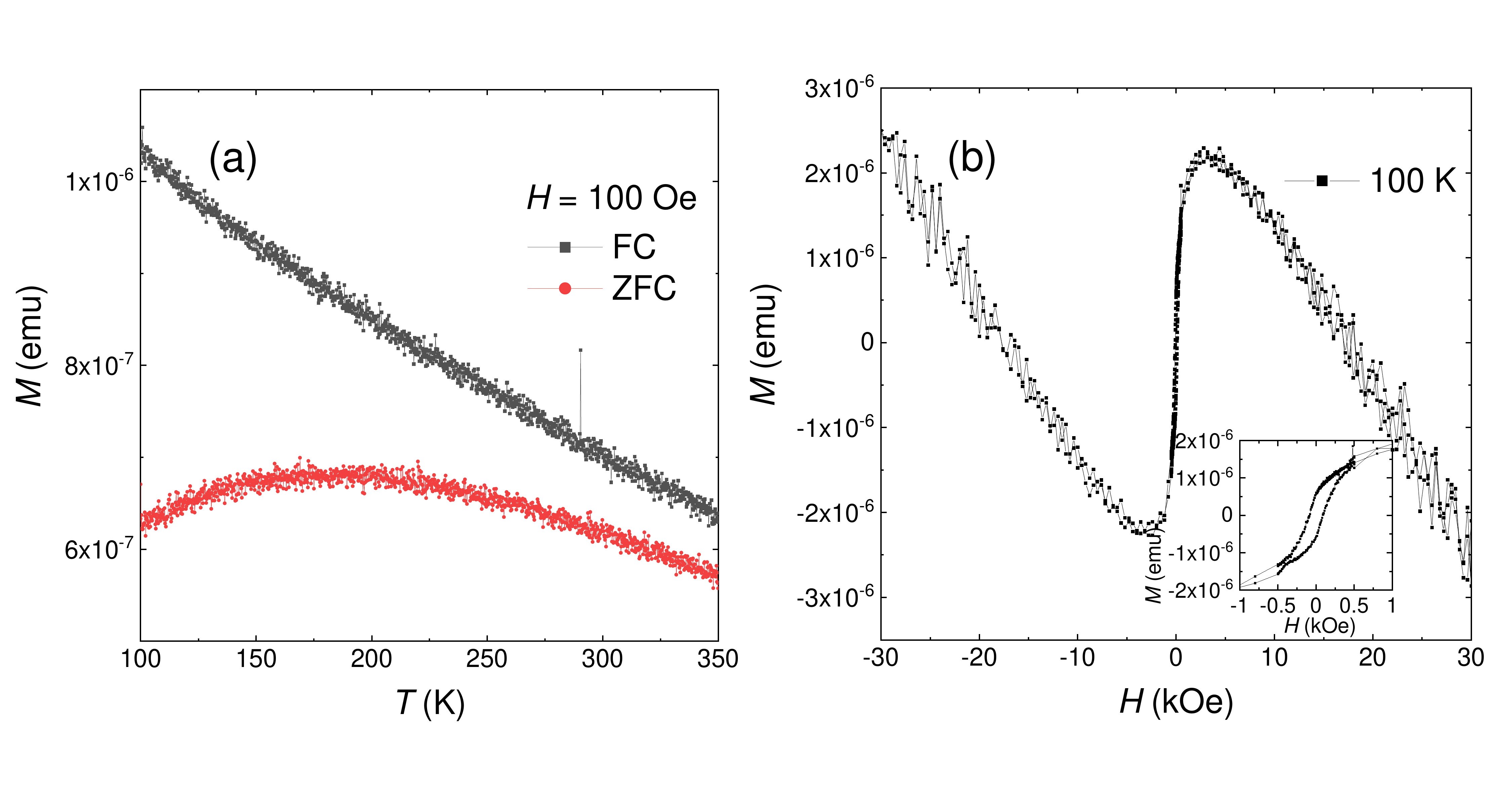}
	\caption{\label{fig4} The magnetization of sample $\mathrm{S2}$. (a)\textit{M-T} in a magnetic field of 100~Oe. (b) Field-dependent magnetization (\textit{M-H}) at 100 and 300~K. Inset: Expanded view of the \textit{M-H} curve ranging from -1~kOe to 1~kOe. }
\end{figure}

Then, we measured the magnetization of a granule sample $\mathrm{S2}$ which began to shake when a $\mathrm{Nd_2Fe_{14}B}$ magnet approached [see Fig.\ref{figs3}]. Since this sample is too small to be accurately weighed, we have directly expressed the vertical axis units in Fig.\ref{fig4} as ``emu". The FC and ZFC measurements of \textit{M-T} curves exhibit similar positive values and similar branching with that of sample $\mathrm{S1}$. This indicates that $\mathrm{S1}$ and $\mathrm{S2}$ have similar magnetic components. 
However, many other samples showed no response to the $\mathrm{Nd_2Fe_{14}B}$ magnet, and some were even smaller than $\mathrm{S2}$. We think that this may be related to the non-uniformity of the samples, when the sample has a suitable size, a suitable composition, and a suitable shape, it is possible to reach the state of half levitation.

\begin{figure}[!htbp]
	\centering
	\includegraphics[width=\linewidth]{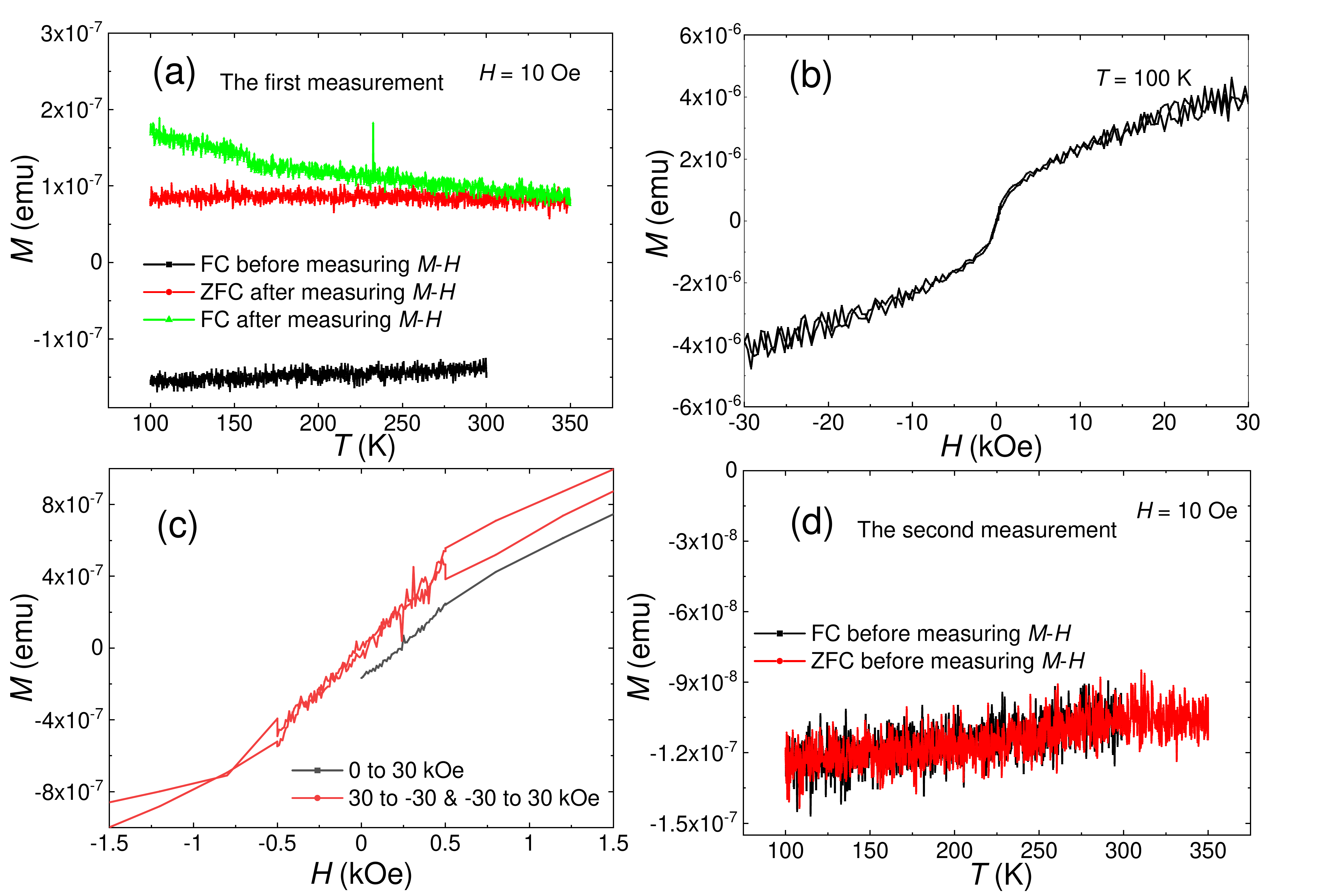}
	\caption{\label{fig5} The magnetization of sample $\mathrm{S3}$. (a) The FC measurement of \textit{M-T} at 10~Oe before measuring \textit{M-H}, denoted as a black curve; the FC and ZFC measurements of \textit{M-T} at 10~Oe after measuring \textit{M-H}, denoted as green and red curves, respectively. (b) \textit{M-H} at 100~K. (c) Expanded view of the \textit{M-H} curve ranging from -1.5~kOe to 1.5~kOe. (d) The second FC and ZFC measurements of \textit{M-T} at 10~Oe.}
\end{figure}

Finally, we measured the magnetization of sample $\mathrm{S3}$ which shows half-magnetic-levitation on a $\mathrm{Nd_2Fe_{14}B}$ magnet. The snapshot of the half-magnetic-levitation of $\mathrm{S3}$ is shown in Supplementary Materials [see Fig.\ref{figs4}]. We first conducted the FC measurements of the \textit{M-T} curve from 100 to 300~K at 10~Oe. The magnetization of the FC curve (black curve) shows a clear negative value in Fig.\ref{fig5}(a), almost unchanged with temperature below 300~K, which is the same as the results in references \cite{Lee01,Lee02,Lee03} and \cite{huake}. 
However, before the ZFC measurements of \textit{M-T}, we conducted measurement of field-dependent magnetization at 100~K, see Fig.\ref{fig5}(b). When the magnetic field increases from 0 to 1500~Oe, the magnetization increases from negative to positive. The black curve in Fig.\ref{fig5}(c) is the zoom-in of this procedure. Unlike samples $\mathrm{S1}$ and $\mathrm{S2}$, when the magnetic field increased above 1500~Oe, the magnetization didn't decrease with the field but increased at a lower slope.
 
We think that in addition to the ferromagnetic components, there are some paramagnetic components in sample $\mathrm{S3}$. Considering the inhomogeneity of the as-grown product, especially on smaller samples, the change of components will have a profound impact on them.
After the measurement of field-dependent magnetization, we performed the FC and ZFC measurements of \textit{M-T} at 10~Oe again and obtained positive values (red and green curves in Fig.\ref{fig5}(a)).
It's important to acknowledge that due to the minuscule magnetic signal from $\mathrm{S3}$, determining the precise signal center could introduce inaccuracies, leading to measurement errors.

To address this, we repositioned $\mathrm{S3}$ by rotating it 180° on the sample rod and subsequently performed a second measurement. This time, we conducted consecutive FC and ZFC magnetization measurements. Both \textit{M-T} curves exhibited negative magnetization at 10~Oe [see Fig.\ref{fig5}(d)]. Notably, the ZFC \textit{M-T} curve displayed a slight inflection around 310~K, consistent with the behavior observed in Lee \textit{et.al.}'s sample $\mathrm{4.1}$ in their figure 3(a) \cite{Lee01}.

\begin{figure}[!htbp]
	\centering
	\includegraphics[width=\linewidth]{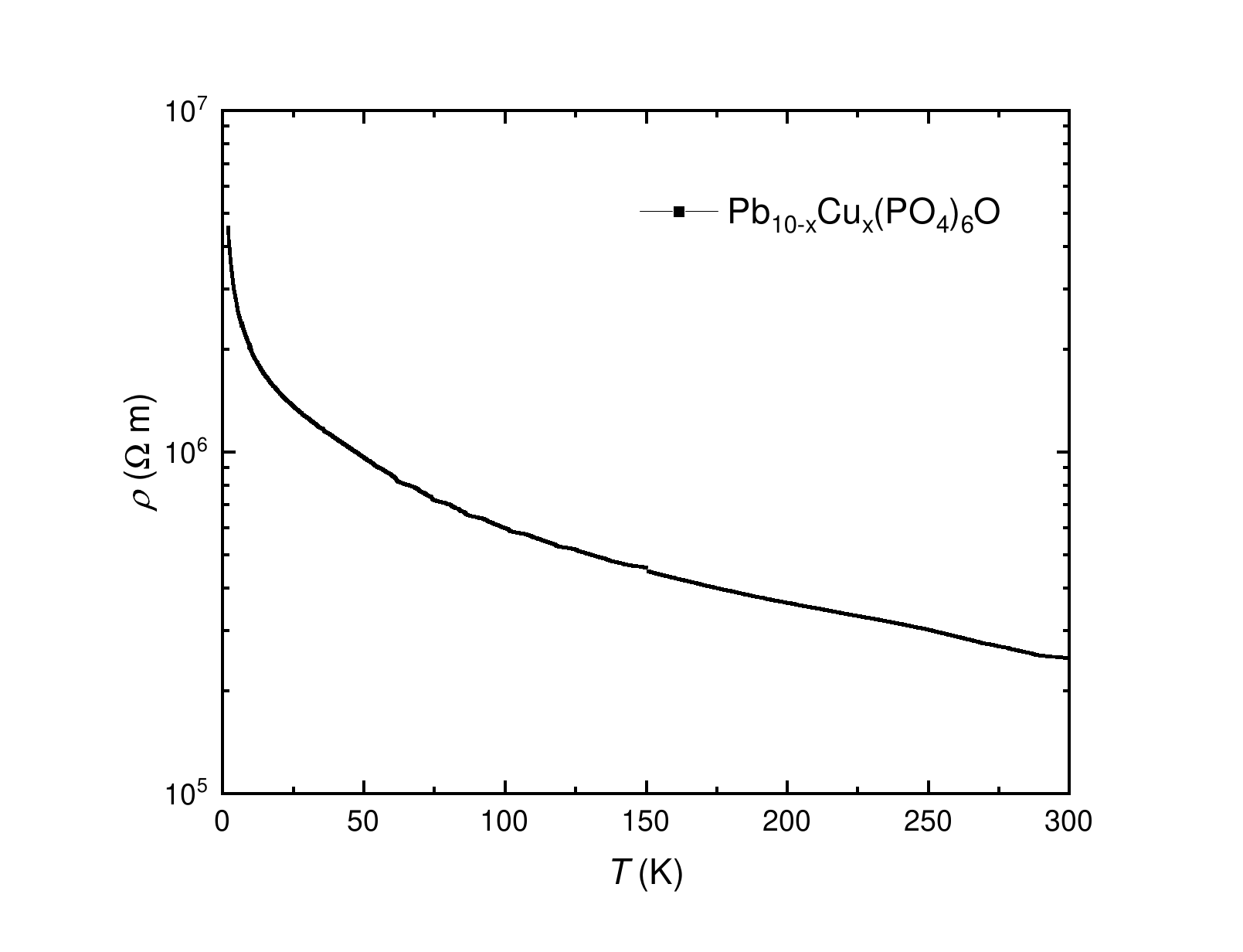}
	\caption{\label{fig6} Temperature-dependent resistivity of our $\mathrm{Pb_{10-x}Cu_x(PO_4)_6O}$ pellet sample.}
\end{figure}

To verify whether our sample has zero-resistivity behavior, we took a resistance measurement to a pellet sample, see Fig.\ref{fig6}. The result shows that our sample has a semiconducting-transport behavior and its resistivity gradually increases as the temperature decreases, from $\mathrm{10^5~\Omega~m}$ improved by an order of magnitude from 300 to 2~K. Some recent repeated experiments have shown similar results to this \cite{beihang,india}.

\section{Analysis and discussion}

To summarize our results: (1) We have synthesized ceramic samples which, despite the inclusion of secondary phases such as Cu$_2$S, are primarily of the same chemical composition and crystal structure as LK-99 samples previously reported \cite{Lee02,Lee03,dongnan}. (2) In terms of magnetic response to external fields, our samples contain at least one linear-diamagnetic phase which is nearly temperature-independent, and one soft-ferromagnetic phase with a Curie temperature above 350 K. The volume fraction of the two phases appears to vary between different sample fragments. (3) The majority phase shows semiconducting resistivity as a function of temperature, and no trace of superconductivity. (4) In our small and flaky sample S3 which exhibits half levitation atop a $\mathrm{Nd_2Fe_{14}B}$ permanent magnet, the magnetic response is dominated by the ferromagnetic phase in the field range ($>1$ kOe) that corresponds to the surface magnetic field of the permanent magnet.

Because magnetic-half levitation in ambient condition might be considered a spectacular and easily accessible phenomenon, it has been a focus of experimental attempts to reproduce the original claim made by Lee \textit{et al} \cite{Lee01,Lee02,Lee03}. Given the existence of both diamagnetic and ferromagnetic phases in our samples, it is important to identify which phase(s) is (are) responsible for the observed half levitation. To do this, we first note the phenomenon of ``shape anisotropy'' associated with a ferromagnetic specimen: it is energetically favorable for the mesoscopic magnetic domains inside the specimen to be aligned along the longest direction of the specimen if the shape is strongly anisotropic (\textit{e.g.}, plate- or rod-like). The energy minimization is associated with the demagnetization field generated by the specimen, hence, the phenomenon would still occur even if the material's magnetization susceptibility is isotropic at the microscopic level. This latter aspect applies to our polycrystalline samples. Importantly, our sample S3 has a highly anisotropic flaky shape, and its half levitation can be viewed as the spontaneous alignment of one of its long axes to the direction of the surface magnetic field of the magnet underneath the sample [Fig.~\ref{figs1}(c)]. The equilibrium is a result of balancing gravitational and magnetic torques exerted on the sample [Fig.~\ref{figs2}].

Our scenario relies on the ferromagnetic property and shape of S3. A distinct expected consequence of it is that there must be a net attraction between the sample and the magnet. In the animation displayed in Fig.~\ref{figs4}, we demonstrate that such attraction indeed exists: the sample can be dragged around (while remaining in a half-levitated state) by careful horizontal movement of the magnet underneath the weighing paper. Furthermore, the sample's motion exhibits distinct discontinuous ``jumps'' when moved, resulting from the static friction between the sample and the supporting paper exceeding the kinetic friction.

We therefore conclude that the half levitation is caused by a magnetic torque, rather than by a net lifting force exerted on the sample. The diamagnetic response of the sample plays no essential role in this phenomenon. As our explanation requires the sample's shape to be anisotropic, it implies that a more significant role of diamagnetism is still possible, if a sample with a more isotropic shape can still be half-levitated. In fact, our sample S2 belongs to that type, but as our animation in Fig.~\ref{figs3} shows, the magnet's movement can only slightly shake the sample rather than levitating it. Finally, we remark that the strength of the ferromagnetism in our sample appears to be strong enough to half-levitate the sample, but \textit{not} strong enough for the sample to be picked up against its own weight by the same magnet from above. However, because the precise chemical composition of the ferromagnetic component remains to be determined, we cannot rule out the possibility that a sample purer of the specific component may be a stronger ferromagnet than our samples. The net attraction between the sample S3 and the magnet could be demonstrated in our dragging experiment only because the friction between the sample and the waxed weighing paper is small.

\section{Conclusions}

To conclude, we have observed a semiconductor-like, non-superconducting electric property in our LK-99-like synthetic samples, along with diamagnetic and soft-ferromagnetic properties arising from supposedly different phases of the mixed product. Our results suggest that one needs to be cautious when interpreting half-levitation observations as evidence for net levitating forces (and, further, as evidence for the Meissner effect). The presence of ferromagnetism in a Pb-Cu-P-O system is somewhat unexpected, as we are not aware of previous reports of materials with related properties. The presence of a flat-band-like electronic structure in $\mathrm{Pb_{10-x}Cu_x(PO_4)_6O}$, as revealed in a recent calculation \cite{griffin2023origin,tavakol2023minimal}, or in $\mathrm{Pb_{9}Cu(PO_4)_6(OH)_2}$, which is believed to form in LK-99 like samples \cite{Princeton}, might be able to give rise to such spontaneous ferromagnetism, which warrants further investigation.

\section{ACKNOWLEDGEMENTS}

We are very grateful to Dr. Xiaoxiao Zhang for sending us 5 grams of lead sulfate as starting material (powder, Aladdin, 99.99$\%$), so that we can synthesize the samples in time.
We wish to thank Professor Victor Galitski from the University of Maryland for helping us upload a high-quality version of supplementary videos (Figs. S3-S4, and more) to ScienceCast and creating an audio description of the content.
This work was supported by the National Key Research and Development Program of China (2021YFA1401900), 
the CAS Interdisciplinary Innovation Team, the strategic Priority Research Program of Chinese Academy of Sciences, Grant No. XDB28000000 and the National Natural Science Foundation of China No. 12141002 No.12225401 and No.U1832214.

\bibliography{rtsc}

\appendix

\section{Supplementary Materials}

\renewcommand{\thefigure}{S\arabic{figure}}
\setcounter{figure}{0} 

\begin{figure*}[!htbp]
	\centering
	\includegraphics[width=\linewidth]{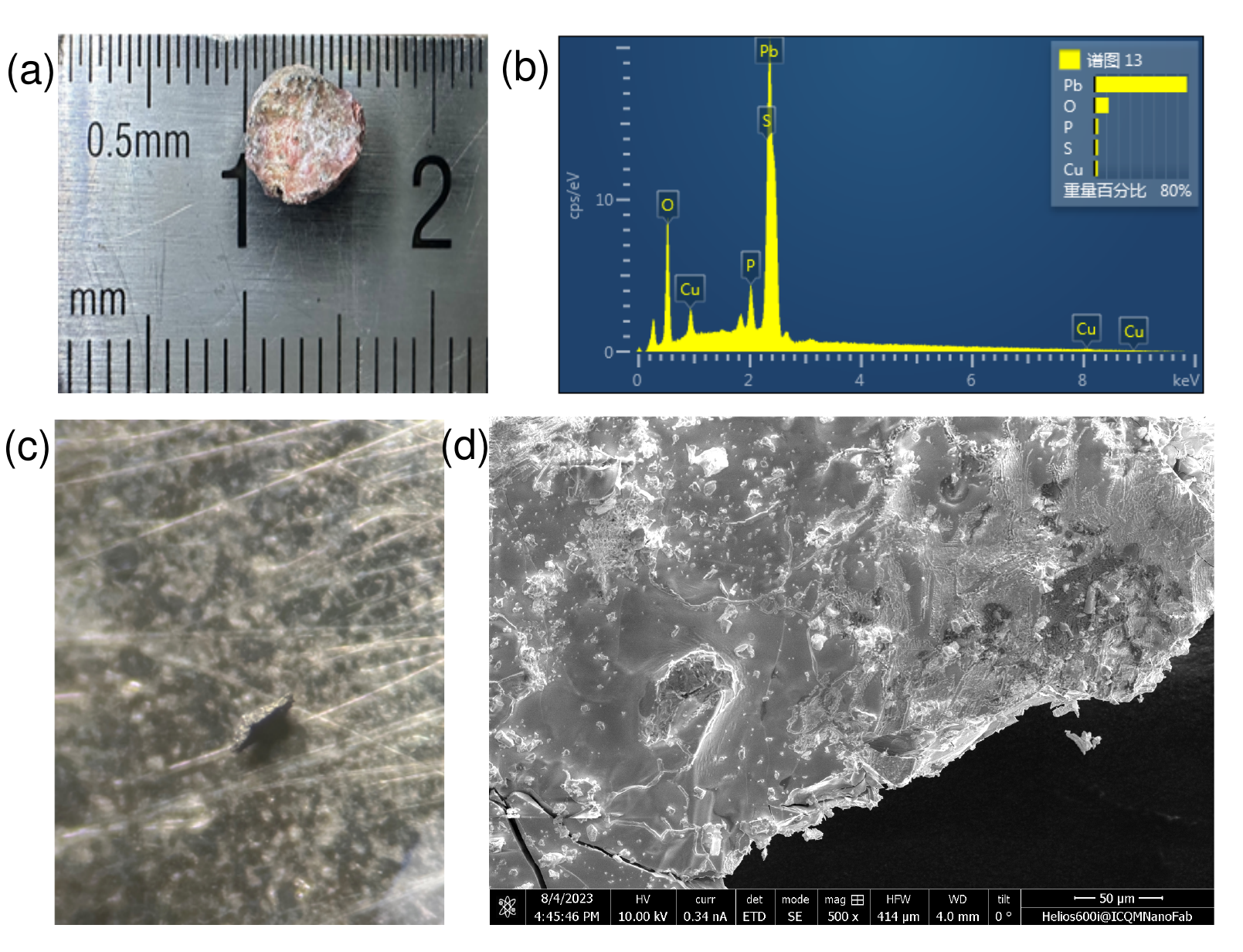}
	\caption{\label{figs1} (a) The as-grown product of $\mathrm{Pb_{10-x}Cu_x(PO_4)_6O}$. (b) The Energy-dispersive X-ray spectroscopy (EDS) of our sample, indicating the presence of lead, phosphorus, copper, oxygen, and sulfur. (c) A photo of half-magnetic-levitated sample. The background is the surface of a $\mathrm{Nd_2Fe_{14}B}$ magnet, captured from a top-down perspective. (d) A scanning electron microscope (SEM) image of a pelleted sample.}. 
\end{figure*}

\begin{figure}[!htbp]
	\centering
	\includegraphics[width=\linewidth]{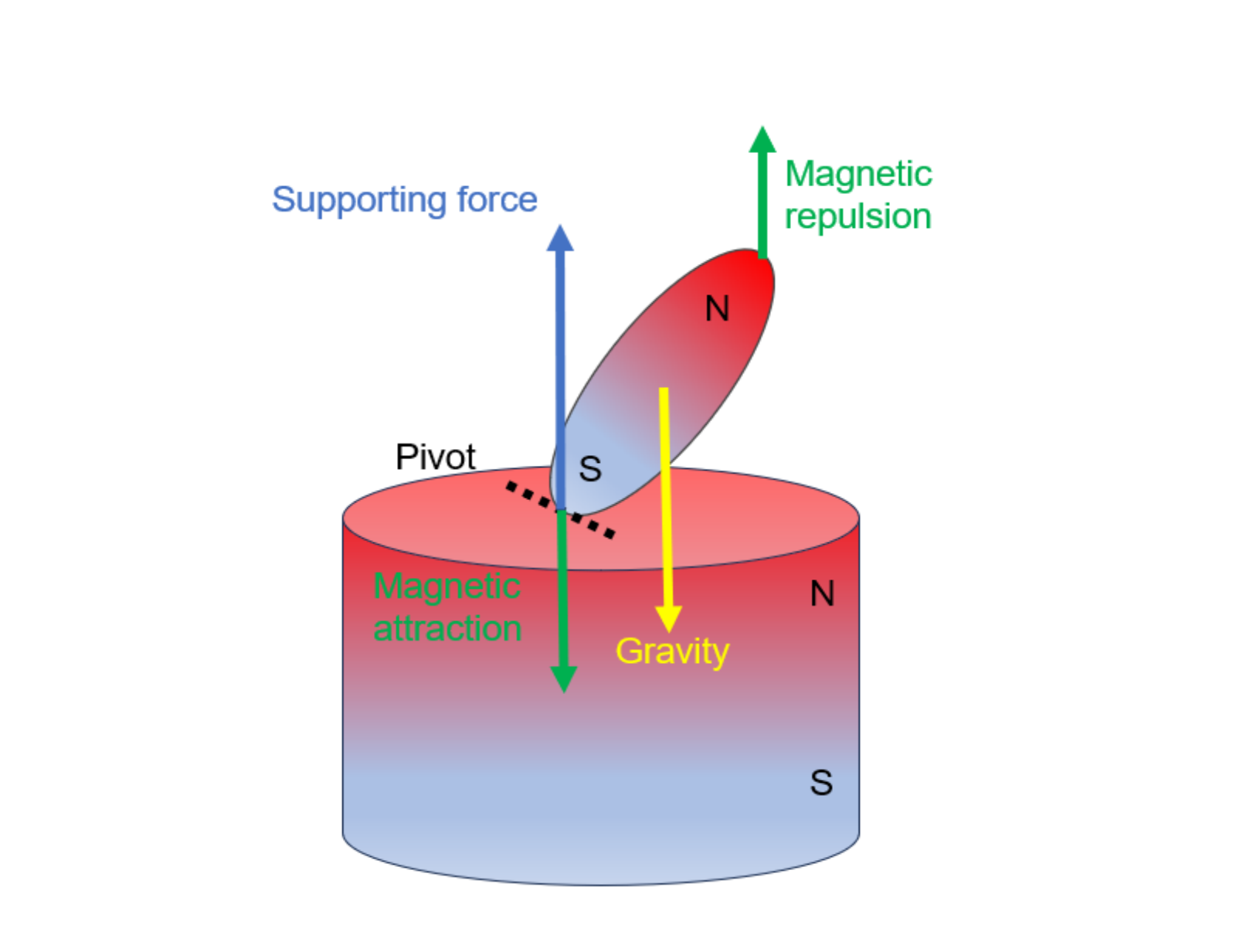}
	\caption{\label{figs2} Schematic diagram of force analysis of a magnetic small piece half-levitated on a magnet.}
\end{figure}

\begin{figure}[ht]
	\centering
	\animategraphics[controls,width=\linewidth]{5}{ssnapshot_}{00}{43} 
	\caption{\label{figs3} Real-time demonstration of sample S2 response to a permanent magnetic underneath the weighing paper. The animation is made of snapshots taken at 0.2 second intervals, and it plays by default at the same speed. The $\mathrm{Nd_2Fe_{14}B}$ magnet under the weighting paper is square-shaped and with its pole facing up towards the viewer. The animation might require designated PDF viewers (such as Acrobat Reader, rather than web browsers) to function normally. The original video can also be viewed online by clicking on ScienceCast in the ``Code, Data and Media'' tab at the bottom of the arXiv page for the present preprint.} 
\end{figure}

\begin{figure}[ht]
	\centering
	\animategraphics[controls,width=\linewidth]{5}{snapshot_}{00}{97} 
	\caption{\label{figs4} Real-time demonstration of half-levitated sample S3 being dragged by a permanent magnetic underneath the weighing paper. The animation is made of snapshots taken at 0.2 second intervals, and it plays by default at the same speed. The $\mathrm{Nd_2Fe_{14}B}$ magnet under the weighting paper is square-shaped and with its pole facing up towards the viewer. The upper and left edges of the magnet are sometimes visible in the animation through the half-transparent weighing paper.
	The animation might require designated PDF viewers (such as Acrobat Reader, rather than web browsers) to function normally. The original video can also be viewed online by clicking on ScienceCast in the ``Code, Data and Media'' tab at the bottom of the arXiv page for the present preprint.} 
\end{figure}

\end{document}